
\font\tsnorm=cmr10

\magnification=1200
\def\aa{{\alpha}}
\def\bb{{\beta}}

\def\dd{{\delta}}

\def\nn{{\nu}}
\def\ee{{\epsilon}}

\def\Om{{\Omega}}
\def\ll{{\lambda}}

\def\gg{{\gamma}}

\def\ref #1{\medskip\item{#1}}
\def\beginsection #1 #2
{\bigskip\bigskip \noindent{\bf #1 \  \  #2} \nobreak\noindent}
\hsize=150truemm
\vsize=8.2truein
\hoffset=6truemm
\voffset=0.25truein
\topskip=25pt
\null
\vskip 5truemm
\vskip 15truemm
\centerline{\bf Static  Solutions of the Einstein Equations}
\centerline{\bf for Spherically Symmetric Elastic Bodies}
\vskip 20truemm

\bigskip\bigskip
\centerline{\tsnorm Jiseong Park
\footnote \dag {Current Address: 134 Ivy Dr. Apt.5, Charlottesville, 
VA 22903, U.S.A.}}
\centerline{\tsnorm park@aei-potsdam.mpg.de}
\centerline{\tsnorm Max-Planck-Institut f\"ur Gravitationsphysik}
\centerline{\tsnorm Albert-Einstein-Institut}
\centerline{\tsnorm D-14473, Potsdam}
\centerline{\tsnorm Germany}

\vskip 0.9in
\centerline{\bf Abstract}
\bigskip

  We analyze the Einstein equations for a spherically
  symmetric static distribution of matter which satisfies
  a given constitutive relation for elastomechanics.  
  After reducing the equations into a system of Fuchsian
  ODE for certain scalar invariants of the strain, we 
  show that,  for a given 
  constitutive relation and a given value of central pressure
  satisfying certain compatibility conditions, 
  there exists a unique regular  solution near the
  center.
  In the case  when  the  constitutive relation
  is given by a quadratic form of strain, we  further show that
  the solutions stay regular up to
  the boundary of the  material ball. 

\vfil
\eject

     
\beginsection \S1 Introduction
\bigskip
It is common in relativistic astrophysics to 
regard  a star as a massive ball consisting of perfect fluid.
Assuming  that the ball is static, the corresponding
Einstein equations reduce to the well known
Tolman-Oppenheimer-Volkoff equation. Various authors 
have discussed  properties of solutions of this equation. The 
 existence 
theorem   for this equation under
general equation of state  has been established
by Rendall and Schmidt  [1]. In the same article,
as well as in a more recent article by Makino [2], the issue of 
finiteness of the ball and the regularity of solutions up to the 
boundary of the ball has been discussed. 

Anisotropic material balls also  have been objects of study and 
many particular solutions of the corresponding Einstein equations have 
been derived (for example [3], [4], [5] and [6])
 under various assumptions imposed on the stress-energy tensor. But,
unlike in the case of the perfect fluid ball, there hasn't been any
general statement concerning the existence of
solutions for anisotropic balls. The purpose of this paper is 
to resolve this problem  for  the
 elastic solids. 
To achieve this purpose, we  
use the relativistic elasticity theory formulated years ago by Kijowski and Magli [7]. 
Although there have been other complete theories of
relativistic elastomechanics 
(for example [8] and [9]), the formalism of
 Kijowski and Magli seems to suit better for our purpose. 

The main idea common in  the modern theories
of relativistic elasticity is that a material property is
chosen by choosing a constitutive relation --- the relation
between the energy stored at a particle and the strain
experienced by that particle. Once a constitutive relation has been chosen, 
the stress can be written as a function of strain and hence the
material variables are  replaced by the dynamic variables
which describe the strain.  
Under this scheme,  the Einstein equations 
for the  spherically symmetric static 
elastic body
can be written  as a closed system of ODE for a 
variable describing the material configuration and two
variables for the spacetime metric. 
In [7] some  families of solutions of these equations 
have been produced, but
 no  theorem regarding the existence
 of solutions of this system under 
general constitutive relations has been given.

In this paper, we show that  
 for a given 
constitutive 
relation  and a prescribed value of central pressure which 
satisfy certain compatibility conditions, there
exists a unique smooth  solution on a neighborhood
of the center to the equations mentioned above. 
To prove the local existence and uniqueness, we use the 
theorem by Rendall and Schmidt [1] which states the 
existence, uniqueness and regularity of solutions for 
a certain  class of Fuchsian ODE systems.

 It is also shown that, in the case the 
given constitutive relation
is a quadratic form of strain,
 the solutions can be extended to 
the boundary (where the radial stress vanishes), so that 
a Schwarzschild vacuum can be continuously joined outside the 
ball. This is done by showing that the solutions are  bounded 
on any finite open interval on which the radial 
stress stays positive. 
The most crucial step  is to show that the mass-radius ratio
is bounded from above by a  constant $< {1\over 2}$. 
The similar issue actually
 arises
in many other matter models under spherically symmetric setting.
 For 
perfect fluid and anisotropic fluid, this estimate
 has been made by Baumgarte and Rendall
 in [10] under the assumption that the tangential stress (in the
case of
anisotropic fluid)
and the equation of state relating the energy density and the 
radial stress have been given and are sufficiently regular. 
In our paper the same estimate is made
but in a different context; the regularity 
of tangential stress is not assumed a priori, but it
is a consequence of the mass-radius ratio estimate. 
  For Vlasov
field, with isotropic or anisotropic stress, the same estimate 
has been established by G. Rein  in [11]. We establish the bound
of mass-radius ratio for the elastic material by adopting the 
method used by Rein in [11].
 
 Throughout the paper, we use the spacetime metric signature
 $(-,+,+,+)$ and the units $c=G=1$. 
\vfill\eject
\beginsection \S2 {A Brief Review of Relativistic Elasticity and }

\noindent{\bf\ \ \ \ \ \
   Generalized Tolman-Oppenheimer-Volkoff  Equations}

\bigskip\bigskip
In relativistic continuum mechanics, it is common to
describe the  configuration of particles  
by  a  
map({\it material map}) from a spacetime into a three dimensional
Riemannian manifold({\it material space}). 
This map $\xi : (M^4, g_{ij})\to (X^3,\gg _{ab})$
must satisfy
the following conditions:
(i) its differential
 $ D\xi$ has full rank everywhere and (ii) at each 
$p\in M$,  the  solution $u^i\in T_p(M)$
of the equation
$u^i\xi^b_{,i}=0, b=1,2,3$ is timelike.
 Both the projection
$\eta _{ij}:=g_{ij} + u_i u_j$ of the  spacetime metric along $u^i$
 and the pull-back $(\xi^* \gamma)_{ij}
:= \xi^a_{,i} \xi^b_{,j} \gamma_{ab}$ of the material metric
 act as  Riemannian metrics on the  subspace of $T_pM$ which is
orthogonal to $u^i$.  
 When $X$
represents  an elastic body, the material metric
$\gamma_{ij}$ describes the infinitesimal distance 
between particles in a locally relaxed state.  So, we say that an
event $p\in M$ is at a relaxed state\footnote
\dag{ But, for a body  under extreme pressure, 
 a relaxed state might not exist. See [8] for a detailed discussion
and an alternative formulation.} 
 if these two tensors
agree on the orthogonal subspace in $T_pM$ of $u^i$.
 There are many different ways of describing the
 deviation of $\eta _{ij}$ from $(
\xi^*\gamma)_{ij}$ and any of them can be called 
 a {\it relativistic strain} tensor. 
 Among them we choose the following definition of strain 
as in [7].

$$S_{i}^{\ j} 
:= -{1\over 2} {\rm log} \Big( (\xi^* \gamma)_{i}^{{\ } j} - u_i
 u^j\Big)\eqno(1)$$
Note that the operator to which ``log'' is applied
is positive and self-adjoint, so the strain is well-defined
by (1) for any choice of $\gamma_{ab}$ and $\xi$. 
One can also show that this tensor  is symmetric and is {\it 
spatial} in the sense that $S_{ij}u^j =0 $.

The main idea in the constitutive theory of  relativistic 
elasticity
 is that
the strain tensor should fully determine the elastic energy $E$   
stored at each particle, that is, $E$ should be a function of $S_{ij}$. 
We call this
function a {\it constitutive relation} for the material. 
Choosing  this
function $E=E(S_{ij})$ is equivalent to choosing a particular type of
material to study. 

The elastic energy $E$ 
as a function of strain determines  
 the 
stress-energy tensor which satisfies the conservation law $ D_i T^i_{\  j} =0$.
 The relationship between the stress-energy tensor and the strain
tensor generalizes the
{\it stress-strain relationship}; the details of its derivation 
can be found in [7]. We will quote the result for an isotropic elastic material only. 
$$T_{ij}=n_o e^{-\aa}\left[E u_i u_j -{{\partial E}\over {\partial \aa}}h_{ij}
-\left({{\partial E}\over {\partial\beta}}\tilde S_{ij}
+ {{\partial E}\over {\partial\theta}}\widetilde{\tilde S
\tilde S}_{ij}\right)\right]\eqno(2)$$
where ``tilde'' means ``traceless part of'', $n_o$ is the 
particle number density measured with respect to the pull-back 
metric $\xi^* \gamma _{ij}$ and
$$\eqalignno
{\aa&:={\rm tr} S &(3)\cr
\bb&:={1\over 2}{\rm tr}[ \tilde S \cdot \tilde S ]&(4)\cr
\theta&:={1\over 3} {\rm tr} [\tilde S \cdot
\tilde S \cdot \tilde S ]&(5)\cr}$$
The quantity $\aa$ measures the
 compression rate of the material  relative to the relaxed state.
More precisely, the particle number density $n$ with respect to 
the projected spatial metric $\eta _{ij}$ is given by 
$$ n= n_o e^{-\alpha}.$$ 
The energy density  is therefore
 $\rho = n E = n_o e^{-\alpha}
E$.
The  part between the round brackets in (2) 
(trace-free part) vanishes identically
if $E$ depends only on the compression $\aa$, 
i.e. when the material is 
a perfect fluid.

\bigskip\bigskip

Now we will consider the case when the spacetime metric $g_{ij}$ is
spherically symmetric and static. We will assume the following form of spacetime metric.
$$ds^2= -e^\nn\ dt^2 + e^\ll\ 
 dr^2 + r^2 d\theta^2 + r^2 {\rm sin}^2\theta d\phi^2
$$
Here the functions $\nn$ and $\ll$ are supposed to be
smooth  on $M$ and  depend only on the radial coordinate $r$. 
This is not the most general form of
spherically symmetric spacetime metric since in general 
$e^{-\lambda}$ can vanish for some $r>0$, but  we will see 
in \S 4 that this cannot happen in the interior 
of material ball.   
We choose   $X=\bf R^3$ as the material space and the  Euclidian metric
\footnote \dag
{For more general material metrics, see  Appendix.}
$$dy^2 + y^2 d\bar \theta^2 + y^2 {\rm sin}^2\bar\theta 
\ d\bar\phi ^2$$
as the material metric.
We also assume that the 
particle distribution at the relaxed state is homogeneous, i.e.
$n_o$ is a constant.
Under this assumption, we may set $n_o\equiv 1$ without loss
of generality. A material map is given by specifying three functions
$y(t,r,\theta, \phi)$, $\bar \theta(t,r,\theta, \phi)$ and
 $\bar \phi(t,r,\theta, \phi)$. But, by 
the assumption that the spacetime is static and spherically symmetric, $y$ 
 must   depend on $r$ only. After adjusting 
the angular coordinates for $X$ if necessary, we may further assume
that $\bar \theta=\theta$ and $\bar \phi=\phi$. So, a material map
is determined by specifying only
 one function $y: [0,\infty)\to [0,\infty)$. We will allow only
smooth material maps, so  $y(r)=rz(r^2)$ for some smooth function
$z:{\bf R}\to {\bf R}$. We also insist  that
$y'(r)>0$  for all $r\in [0,\infty)$
in order to ensure that the material map has full rank everywhere.

With a  function $y(r)$ given for the material map, 
we can now form the strain tensor from (1). We have
$$S=-{\rm log} {{y'(r)}\over {\sqrt{e^{\ll(r)}}}}\left( e^{\ll(r)}dr^2\right)
-{\rm log} {{y(r)}\over r}\left( r^2 d\theta^2\right)
-{\rm log} {{y(r)}\over r}\left( r^2 {\rm sin}^2 \theta d\phi^2\right)
\eqno(6)$$
To write the stress-strain relationship (2), we need
the partial derivatives of the energy with respect to the
invariants of the strain. 
As a consequence of spherical symmetry, there are only two
independent invariants of the strain tensor. Hence, the elastic energy $E$
can be considered as a function of these two independent invariants
of the strain. 
It turns out that the following choices of two invariants make the 
expression of the stress-energy tensor much simpler.

$$u:= {\rm log} \left({{r y'}\over {y\sqrt {e^\ll}}}\right)\eqno(7)$$
$$v:= \aa + u = -3 {\rm log} \left( {y\over r} \right)\eqno(8)$$
The  invariants $\aa,\bb$, and $\theta$
 are related to $u$ and $v$
as follows. 
\def\Eu{{
{{\partial E}\over {\partial u}}
}}
\def\Ev{{
{{\partial E}\over {\partial v}}
}}
$$\aa= v-u, \ \ \ \ \ \ \  \bb= {1\over 3} u^2\ \ \ \  \ \ \ \theta=
- {2\over {27}} u^3 \eqno(9) $$
Using (2) and (9), we now get the stress-energy tensor.
$$T_{ij}= e^{u-v}\left[ 
E e^\nn dt^2  + \Eu e^\ll dr^2 +\left(\Eu - {3\over 2}(\Eu +\Ev)\right)
r^2(d\theta^2+{\rm sin}^2\theta d\phi^2) \right]
\eqno(10)
$$
\def\OO{{\Omega}}

\bigskip\bigskip
Let us pick  notations for  the components of the stress-energy tensor.
$$\eqalignno{ \rho &:= e^{u-v} E &(12)\cr
 P&:= e^{u-v} \Eu &(13)\cr
 \OO &:={1\over 2} e^{u-v}(\Eu +\Ev) &(14)\cr
Q&:= e^{u-v} \left(\Eu - {3\over 2}(\Eu +\Ev)\right) &(15)\cr}
$$
The Einstein equations can be now written.
$$\eqalignno
{8\pi r^2\rho&=e^{-\ll}(r\ll'-1)+1 & (16)\cr
 8\pi r^2 P   &=e^{-\ll}(r\nu'+1)-1 &(17)\cr
 8\pi\  Q      &={1\over 2} e^{-\ll}\left( \nu''+{1\over 2}\nu'^2
                 +{1\over r}(\nu'-\ll')-{1\over 2}\nu'\ll'\right)&(18)
 \cr
}
$$
It should be remarked at this point  that this is a closed system for three
unknowns $\{y, \ll, \nn\}$ since $E$ is a function of $u$ and $v$,
which are again functions of $y, y'$, and $e^\ll$. 

\bigskip\bigskip \ \ 
Next,  we will modify the Einstein equations 
(16)-(18) into the  generalized 

\noindent
 Tolman-Oppenheimer-Volkoff equations by introducing a new variable
$$w:= {1\over {2r^2}}\left( 1- e^{-\ll}\right) \eqno (19)$$
With this new variable, the equation (16) becomes
$$rw' = 4\pi \rho - 3w \eqno(20)$$
The other two equations become
$$P' = - {6\over r}\Om -r (1-2r^2w)^{-1}\left(4\pi P+w\right)\left( P+\rho
\right)  \eqno(21)$$
and
$$\nu ' = 2r (1-2r^2 w)^{-1}(4\pi P +w) \eqno(22)$$
The equation (22) can be integrated after all the other solutions
have been found. So, we will focus on the equations (20) and (21) only.

We will assume throughout the
paper that the constitutive function $E(u,v)$ is smooth. We will be also
interested in only smooth
solutions, that is, smooth material map and smooth 
metric coefficients. With these assumptions, it follows 
 from (12)-(15) that all the variables which 
appear in the equations have to be smooth. 
So, we may consider them as smooth functions of $x=r^2$ defined on $\bf R$.
The equations (20)-(21) then can  be expressed using the new independent
variable $x=r^2$ as follows. 
$$\eqalignno
{ 2xP' &= -6\OO - x(1-2xw)^{-1}(4\pi P + w) (P+\rho) & (23)\cr
  2xw' &= 4\pi \rho - 3w & (24)\cr
}$$
Here, and throughout the rest of the paper,
 the ``prime'' indicates the derivation
with respect to $x$.
We also have an auxiliary equation
from (7) and (8),
$$2xv' = 3-3(1-2xw)^{-{1\over 2}} e^u \eqno (25)$$

The equations (23)-(25)
form a closed system for the variables $u$, $w$, and $v$. If a set of smooth
solutions $\{u(x), w(x), v(x) \}$   are found, then the solutions
for the original equations will be determined algebraically or by a 
simple integration. The function
$y$ will be determined from  (8) since it can be explicitly
solved for $y$ as $y(r)= r e^{-v(r^2)}$. The metric variables
$\ll$ and $\nn$ will be determined by   (19) and (22) respectively
\footnote\dag
{(19) can be solved for $\ll$ only if we know that $w<(2r^2)^{-1}$. It
will be shown in \S 4 that indeed this inequality holds on any interval
on which the radial stress $P$ never vanishes. } 
 

\vfill\eject

\beginsection \S3 {Local Existence and Uniqueness of Solutions }

\bigskip

This section is concerned with the integrability of the 
equations (23)-(25)
on a neighborhood of $x=0$. 
After suitable substitutions  and rearrangements 
we  can write the equations explicitly
in terms of $u, w$ and $v$.
$$\eqalignno
{\hskip -3 pt 2xu'  
&= \hskip -3 pt
{   {3(E_u-E_{uv})(1\hskip -3 pt -\sqrt b e^u)\hskip -3 pt -3(E_u+E_v)}
\over
     { (E_u+E_{uu})}                }
-{  {xb(4\pi e^{u-v}E_u+\hskip -3 pt w)(E\hskip -3 pt +\hskip -3 pt E_u)}
\over
      {E_u+E_{uu}}               } & (26)\cr
2xw'&= 4\pi e^{u-v} E - 3w & (27) \cr
2xv'& = 3 - 3\sqrt b e^u &(28)\cr
}$$
where 
$$b:= (1-2xw)^{-1} \eqno (29)$$
and $E$ is a smooth function of $u$ and $v$. These equations are
singular at $x=0$ and therefore the existence theorem
for regular equations cannot be applied here. But, there exists a
theorem which states  existence, regularity and uniqueness of solutions
for
this type of equations, which we state below.

\bigskip\bigskip
 
\noindent{\bf Theorem 1 (Rendall and Schmidt).} {\it
 Let $V$ be a finite dimensional
vector space, $N: V\to V$ a linear map all of whose eigenvalues have 
positive real parts, and $G:V\times(-\ee, \ee) \to V$ and 
$g:(-\ee,\ee)\to V$ smooth maps, where $\ee>0$. Then, there exists $\dd<\ee$
and a unique bounded $C^1$ function $f:(-\dd,0)\cup(0,\dd)\to V$ which 
satisfies the equations
}
$$x{{df}\over{dx}} + Nf = xG(x, f(x)) + g(x)\eqno (30)$$
{\it Moreover, $f$ extends to a smooth solution of } (30) {\it
on $(-\dd,\dd)$. If $N$, $G$ and $g$ depend smoothly on a parameter $t$
and the eigenvalues of $N$ are distinct, then the solution 
depends smoothly on $t$.}

\bigskip

Our equations are not in the form (30) yet.
For instance, the equation (26)  contains  terms
which  are non-linear
in  $u$ and $v$ and those terms do not have a factor of $x$.
One way to introduce a factor of $x$  in those terms is to write
$u= u_o + xu_1$ and $v=v_o+xv_1$ and to rewrite the equations for
the new variables
$u_1$, $w$ and $v_1$.
By the way, (28) implies that
  $v(0)$ can be chosen freely and
 $u(0)=0$ necessarily. 
So, we let 
$$
u= xu_1 ,\ \ \ \ \ \ \ \ \ \ \ \ \ \ \ \ \ \ \ \ \ 
v= v_o + xv_1 $$
Then the equations (26)-(28) become 
$$\eqalignno
{2xu_1'+2u_1  
&=
{   {3(E_u-E_{uv})(1-\sqrt b e^u)-3(E_u+E_v)}
\over
     {x (E_u+E_{uu})}                }&\cr
&\ \ \ -{  {b(4\pi e^{u-v}E_u+w)(E+E_u)}
\over
      {E_u+E_{uu}}               } & (31)\cr
2xw'+3w \hskip 1 pt  &= 4\pi e^{u-v} E  & (32) \cr
2xv_1'+2 v_1&= 3x^{-1}(1 - \sqrt b e^u) &(33)\cr
}$$
We will now show that, if
$E(u,v)$ and $v_o$ satisfy  certain conditions, then these equations can be 
rearranged
into a system of the form given  in  Theorem 1.
Let $U:=(u_1, w, v_1)$.  

\bigskip

\noindent{\bf Lemma 1.} {\it Suppose (i)
$E_u(0,v_o)+ E_{uu}(0, v_o)\ne 0$ and (ii)$ E_u(0,v) + E_v (0, v)
=0$ for all $v$. Then, the 
equations {\rm  (31)-(33)} can be rearranged into
the form}
$$2xU'+\Lambda U = x{{\cal G}(U,x )} + V \eqno(34)$$
{\it Here ${\cal G } : {\bf R^3}\times {\bf R}\to {\bf R^3} $
 is a smooth map, $V$ is a 
constant vector in $\bf R^3$ and }
$$\Lambda=\pmatrix
{ 5
& 
k
&0\cr
{} &{} &{}\cr
0 &3 &0\cr
{} &{} &{}\cr
3 &3 &2\cr
}_,$$
{where $k$ is a constant which depends on $v_0$.} 

\bigskip

\noindent{\bf Proof.} 
Throughout the proof,  ``$O(x)$'' will represent a generic function
of the form $x f(u_1,w,v_1,x)$ where $f$ is  a function which is
smooth on a neighborhood of the set $x=0$ in $\bf R^4$. 

\medskip\noindent
(1) {\bf The First Equation:}  The first term 
on the right hand side of (31) is
$$3( {E_u+E_{uu}})^{-1}
\left[ (E_u-E_{uv})x^{-1}\left({{1-\sqrt b e^u}}\right) - x^{-1}({E_u+E_v}
)
\right]$$
We have
$$\eqalignno{x^{-1}
({1-\sqrt b e^u})& = {x}^{-1} \left[ 1- (1-2xw)^{-{1\over 2}}
e^{xu_1}\right]&\cr &={x}^{-1}\left[1- (1+xw)(1+xu_1) + O(x^2)\right]&\cr
& = -w - u_1 +O(x)&(35)\cr}
$$
and
$$(E_u - E_{uv})(u,v) = (E_u - E_{uv})(0, v_o) + O(u)
=  (E_u - E_{uv})(0, v_o) + O(x)$$
So,
$$(E_u - E_{uv})x^{-1}
({1-\sqrt b e^u}) = (E_u - E_{uv})(0, v_o) \cdot ( -w - u_1) +O(x)
\eqno(36)$$
Next, using the second assumption, $(E_u+E_v)(0,v)\equiv 0$, we have
$$\eqalign{(E_u+E_v)(u,v)&=xu_1(E_u+E_v)_u(0,v_o)+xv_1 
 (E_u+E_v)_v(0,v_o)+ O(x^2)\cr
&=xu_1(E_{uu}+E_{uv})(0,v_o) + O(x^2)\cr
}$$
So,
$${x}^{-1} (E_u+E_v) (u, v) =
u_1\cdot (E_{uu}+E_{uv})(0, v_0) + O(x)\eqno(37)$$
Combining (36) and (37), we get
$$\eqalign {
 (E_u-E_{uv})x^{-1}&\left({{1-\sqrt b e^u}}\right) - x^{-1}({E_u+E_v}
)\cr
&= (E_u-E_{uv})(0,v_o)\cdot (-u_1-w) - (E_{uu}+E_{uv})(0,v_o)
\cdot u_1 + O(x)\cr
& =- (E_u+E_{uu})(0, v_0)\cdot u_1 - (E_u-E_{uv})(0, v_0)\cdot w + O(x)\cr
}$$
On the other hand, by the first assumption, $(E_u+E_{uu})^{-1}$ is
a smooth at $(0, v_o)$. So, $(E_u+E_{uu})^{-1} = (E_u (0, v_o)
+E_{uu}(0, v_o))^{-1} + O(x)$.
Therefore,
$${   {3(E_u-E_{uv})(1-\sqrt b e^u)-3(E_u+E_v)}
\over
     {x (E_u+E_{uu})}                }= -3 u_1 - 
3\left({ {(E_u-E_{uv})(0,v_0)}\over {(E_u+E_{uu})(0,v_0)}
}\right)w + O(x)\eqno(38)$$

The second term on the right hand side of (31)
is of the form $b\cdot g(u,w,v)$, where $g$ is a smooth function
which is 
 linear in $w$. 
Since $b = (1-2xw)^{-1} = 1+O(x)$  we have
$$\eqalignno
{b&\cdot g(u,w,v) &\cr
&= g(0,0, v_o)+xu_1 g_u (0,0, v_o) + w g_w(0,0, v_o) + xv_1 g_v(0,0, v_o)
+ O(x) &\cr
&=C_1 +C_2 w + O(x)& (39) \cr}$$
for some constants $C_1$ and $C_2$ which depend on $v_0$.
Therefore, adding  (38) with (39), the equation (31) becomes
$$2xu_1' + 2u_1 = -3 u_1 -k w + O(x) + V_1$$
or
$$2xu_1' + 5u_1 +kw = O(x) + V_1$$
for some constant $V_1$.

\bigskip

\noindent (2) {\bf  Second and Third Equations:}
Since $e^{u-v} E(u,v)$ is smooth, we can write
this as $C_1 + xu_1 C_2 + xv_1 C_3 + O(x)= C+O(x) $.  So, 
the second equation is already in the desired form.
The third equation has been already handled by
(35).

\bigskip

\noindent{\bf [End of the Proof of Lemma 1]}

\bigskip
\bigskip

\noindent{\bf Remark:}
The first hypothesis of this lemma
restricts the  choice of data $v_o$ when a constitutive relation is given. 
But,unless $E_u+E_{uu}\equiv 0$ on the line $u=0$, 
there will be always an open interval $I$ such that the hypothesis is
satisfied for all $v_o\in I$. This tells us that for almost all
the choices of constitutive relation, there will be enough room to
pick up $v_0$ with which the first hypothesis is satisfied. 

\medskip
\bigskip\bigskip

A local existence theorem immediately follows
from Theorem 1 and the above lemma.

\bigskip
\noindent{\bf Theorem 2.} 
{\it Assume  $E(u,v)$  and $v_o$
 satisfy the hypotheses of Lemma 1.}
{\it Then there exists a  positive constant $\dd$ 
 such that   the equations} (27)-(28) {\it admit a unique set
of $C^\infty$ solutions  $\{u, w, v \}$   on the interval
 $(-\dd, \dd)$  satisfying
$v(0)=v_o$.
}
\bigskip\bigskip

Suppose that the constitutive relation is given in the 
 following  form 
$$E = m + {A\over 2}(u-v)^2 + {B\over 3} u^2  \eqno(40)$$
where $m$ represents the 
rest mass energy of a particle and  $A$, $B$ are nonnegative 
constants.
Then, 
 the condition
(i) is satisfied for all $v_o $ except for $v_o = 1+ {2B\over{3 A}}$
and the condition (ii) is also satisfied.

\vfill\eject

\beginsection \S4 {Global Properties of Solutions for Quasi-Linear Models}

\bigskip

In this section we study the global properties
of solutions under the assumption that the constitutive relation is
given by (40).  
The elasticity theory under such a form of constitutive relation
is called the {\it quasi-linear elasticity} because
under this constitutive relation the stress-strain relationship
becomes linear.  Under the quasi-linear constitutive relation of the 
form (40), the stress-energy tensor components
(12)-(15) becomes 
$$\eqalignno{ \rho &:= e^{u-v} \left[m+ {A\over 2} (u-v)^2 + {B\over 3} u^2
\right] &(41)\cr
 P&:= e^{u-v}\left[A(u-v) + {2B\over 3} u\right] &(42)\cr
 \OO &:={B\over 3}u e^{u-v} &(43)\cr
Q&:= e^{u-v} \left[A(u-v) - {B\over 3} u\right] &(44)\cr}
$$

One of the properties of solutions in  which we are interested is whether
the radial stress $P$ vanishes at some finite radial distance or not.
If $P$ vanishes at some finite radial distance $r=R$, we can consider
the sphere $r=R$ as the boundary of the material body. If, further, the
spacetime metric stays bounded inside this sphere, then
a vacuum Schwarzschild solution may be attached outside the sphere
to obtain a complete $C^1$ spacetime. Therefore another important question 
is whether the solution stays finite up to the radial distance 
where the radial stress vanishes. 

This section focuses on answering the last question, but the first question
will be answered  in a special case 
when the anisotropy constant $B$ is sufficiently close to $0$.
From now on
we  assume that $v(0)=v_o<0$ is given. Since $u(0)=0$, this means that
we are assuming that the radial stress $P$ at $x=0$ is positive.
We will show below that if $P$ stays positive on a finite
interval then all the variables $u$, $v$ and $w$ stay bounded
and the equations (26)-(28) stay regular (except at $x=0$) on the interval. 
The standard extension theorem of ODE then would imply that the
interval of existence can be enlarged, i.e. the interval is not
the maximal interval of existence yet.

\bigskip\bigskip

\noindent{\bf Bounding $w$.}

\bigskip

Let $[0,R)$, $R<\infty$,
 be an interval of existence of  regular solutions of our equations 
(23)-(25) and assume $P>0$ on the interval. Suppose $w$ were not bounded from
above
on the interval. Then by continuity of $w$ on the interval, it would
 follow that
$1-2x_1 w(x_1) =0$ for some $x_1\in [0, R)$. Then from the equation (23),
$\OO$ would have to have a singularity at $x=x_1$. This contradicts the 
assumption that the  solution is regular on $[0, R)$. Therefore 
$w$ must be bounded from above. 
To obtain the lower bound, we use the equation (20) from which it follows 
 that $w(0)= 4\pi \rho (0)>0$ and  
$${d\over {dr}}(r^3w(r))= 4\pi \rho r^2 \eqno(45) $$
 for all $r$. So, $r^3w(r)\ge 0$
for all $r$. Therefore,
$w\ge 0$. Moreover, because ${d\over {dr}}(r^3w)$ is strictly positive
(since $\rho>0$),
$w$ must be bounded from below by a positive constant if we restrict
the interval to $[x_o, R)$ for some $x_o>0$.

\bigskip\bigskip

\noindent{\bf Bounding $\bf e^\ll$.}   

\bigskip

First we will  check whether the integral of $\rho$ is bounded
on finite intervals. From (45),
we have $4\pi \rho (r) = r^{-2} {d\over {dr}}(r^3 w(r))$.
Integrating this, we get
$$4\pi \int_0^r \rho (s) ds 
= r w(r) + 2 \int _0 ^ r w(s) ds \eqno (46)$$
So, it follows that the integral of $\rho$ must be bounded on any 
finite interval since we know that  $w$ is bounded. 
Now, we are ready to show that $e^\ll$ is bounded. The proof of following
lemma is based on an  argument that has been used by G. Rein in [9].

\bigskip
\noindent{\bf Lemma 2.} {\it  Let
 $[0, R)$ be a finite interval on which the  regular solution exists
and $P>0$.
Then, $e^{\ll} $ is bounded on the interval.}

\bigskip
\noindent{\bf Proof.} 
Using the original form of the Einstein equations (16) and (17)
and the generalized Tolman-Oppenheimer-Volkoff equation (21)
we can derive  the following
equation.  
$$\left(e^{{\ll+\nn}\over 2} (4\pi P +w)\right)' 
= 4\pi e^{{\ll+\nn}\over 2} \left({\rho\over {2x}}
-{{3w}\over {8\pi }} - {{3\OO}\over x}\right) \eqno (47)$$
On the other hand,
from (41)-(44), it follows that for a choice
of constant 
$$k > {\rm max} \left( \sqrt {A\over m} , 
\sqrt {3B\over {2m}}\right)$$
we have
$$ |P|< k\rho \ \ \ \ \ \  {\rm and} \ \ \ \ \ \ \ \  |Q|< k\rho$$
and hence
$$|\Omega | \le 2k\rho$$
So, noting that $w\ge 0$, we get  the following 
inequality  from the equation (47).
$$\left(e^{{\ll+\nu}\over 2}(4\pi P +w)\right)^{\prime}
\le K \rho \left(e^{{\ll+\nu}\over 2}\right)$$
for some positive constant $K$. By restricting the
interval to $[x_0, R)$ for some $x_o>0$ if necessary, we may assume that
$w$ is bounded from below by a positive constant. Then, the above
inequality can be replaced by  
$$\left(e^{{\ll+\nu}\over 2}(4\pi P +w)\right)^{\prime}
\le K \rho \left(e^{{\ll+\nu}\over 2}(4\pi P +w)\right)$$
Applying Gronwall type argument to this inequality, we get
$$\left(e^{{\ll+\nu}\over 2}(4\pi P +w)\right)(x) \le 
C {\rm exp } \left( K \int_{0}^x \rho(s) ds \right) $$
for some positive constant $C$. 
So, using  (46) and the earlier 
observation that $w$ is bounded, we get
$$ e^{{\ll+\nu}\over 2}(4\pi P +w) \le C $$
on $[0, R)$.
From here, using again the fact that $w$ is bounded from below by a 
positive constant we get
$e^{{\ll+\nn} \over 2} \le C $.
Now, the desired bound
$e^{\ll\over 2} \le C$ follows since by the equation (22)
$\nn$ is monotonically increasing.

\noindent {\bf[End of Proof]}

\bigskip

\bigskip
\noindent{\bf Bounding the Rest.}

\bigskip

As before, we assume that $P>0$ on the interval $[0, R)$.
\medskip
\noindent{\bf Step I.}
{\it Showing that  $v$, $P$ and $\OO$ are bounded from above.}
It follows immediately from  the equation (25) that $v$
is bounded from above. 
We get from (23)  
$$2 (xP)'= 2xP' - 2P  \le -6\OO- 2P  \le 6|\OO| + 2|P|
\le k\int_0^x\rho(s)ds$$  
where, in the last step, we used the fact that $|P|$ and $|\OO|$
are bounded by $k\rho$ for some constant $k$. 
This gives an upper bound of   $P$ since we know that
the integral of $\rho$ is bounded.
The upper bound of $\OO$ then follows from
$$2\OO = P - A(u-v) e^{u-v} \le P +1$$

\noindent{\bf Step II.}  {\it Showing that $u$ is bounded from above.} 
If $B>0$, we  have
$$ue^u = {3\over B} \OO e^v\le C $$
So, $u$ is bounded from above.
 If $B=0$, then 
we know that  $A>0$, so we have 
$$(u-v)e^{u-v} = {P\over A}\le C $$
Therefore $u-v$ is bounded from above. 
Since $v$ is already
bounded from above, it follows that $u$ is bounded from above.

\medskip
\noindent{\bf Step III.} {\it Showing that $u$ and $v$ are bounded from below
and  the equation {\rm (26)} remains regular within the 
interval $[0, R)$}. 
A lower bound of $v$ comes  from the equation (28)
since  we know that $\sqrt b:= e^{\ll\over 2}$ is bounded and that 
$u$ is bounded from above. 
 $u$ is bounded from below since 
$$e^{v-u} P = \left(A+ {2B\over 3}\right) u - Av >0$$
and $v$ is bounded from below.
The singularity of the equation (26) outside $x=0$ can occur only when 
$E_u+E_{uu}$ vanishes. But, from (40),
$$ E_u +E_{uu} = A(u-v) + {2B\over 3} u + A+ {2B\over 3}
= e^{v-u} P + A + {2B\over 3} > A + {2B\over 3} >0$$
So, there is no singularity away from $x=0$.

\bigskip\bigskip
We have shown so far

\bigskip

\noindent
{\bf Theorem 3.} {\it 
Let  $E(u,v)$ be  given by {\rm (40)}.
 Given  a constant $v_0<0$, 
let $[0,R)$ be 
the maximal interval on which  the regular  solution of 
{\rm (26)-(28)} satisfying the initial condition  $v(0)=v_o$ exists. 
Then $P$ must vanish in the interval unless $R=\infty$.}
 
\bigskip\noindent
{\bf Remark: } Note that this theorem applies only to the quasi-linear
constitutive relation while the local result (Theorem 2)
could be obtained for much larger class of constitutive 
relations. Many of the arguments  used to  prove Theorem 3  rely
on special features of the quasi-linear constitutive relation 
given by (40). But, the most crucial step --- the  boundedness 
of  $e^\ll$  --- could
have been established  for a much larger family of constitutive relations,
namely those which satisfy the {\it dominant energy condition}.  
Note that under the dominant energy condition, the quantity
$|\Omega|$ is bounded by $2\rho$, which was the most important
step in the proof of Lemma 2. 
 
\bigskip\bigskip\bigskip
\noindent{\bf Finiteness of Radius}

\bigskip

 Suppose $B=0$. Then $\Om=0$, so the material is an isotropic fluid. We will show
that the radius is finite. Suppose $P>0$  for all $x$. Then, by the 
monotonicity of $P$ (see equation (23)), we know that
 $P$ tends to a limit as $x\to \infty$.
So, $P'$ tends to $0$ as $x\to \infty$. Then, again by the equation (23), it follows that
$\rho \to 0$ and $P\to 0$ as $x\to \infty$.  But
$$ \rho = e^{u-v} \left (m + {A\over 2} (u-v) ^2\right)\hskip 0.5 in  {\rm and} \hskip 0.5in
P= A e^{u-v} (u-v).$$
We can see from here  that it is impossible for $\rho$ and $P$ both tend to $0$ while $P>0$.
Therefore $P$ must vanish somewhere. 
By the continuous dependence of solutions  on parameters 
 (Theorem 1), it immediately follows that for all sufficiently small $B$,
the solutions must represent balls with finite radius. 






\beginsection \S5 {Discussion}
\bigskip

We have shown that, given a  smooth constitutive
relation and a prescribed value of the stress at the center,
 there exists a unique static
 relativistic elastic ball. We also have shown, if
the elastic energy is  quadratic in the strain tensor,
then  either  the radial stress vanishes at a finite radial distance
before the solution becomes singular
or the material fills up the whole space.  If the 
radial stress vanishes at a finite radial distance $r=R$, then we
can join a Schwarzschild vacuum outside the sphere $r=R$ and obtain
a complete asymptotically flat 
$C^1$ solution of the Einstein equations.
In the case that the radial stress never vanishes, which is certainly
possible in theory, 
we can ask whether the solution represents 
an asymptotically flat spacetime or not, 
although it is hard to imagine an asymptotically flat spacetime
filled with elastic material. 
In any case, it would be nice to be able to tell, by looking at 
a given constitutive relation, whether the solution under that constitutive
assumption represents an elastic ball of finite radius or infinite
radius.

\bigskip
\bigskip

\noindent {\bf Acknowledgments}

\bigskip

The author is very grateful to Alan Rendall  for numerous 
suggestions and remarks. 
 The author also thanks Professor Giulio Magli and Professor Bernd Schmidt
for many useful and important comments. 

\vfill\eject

\centerline{\bf Appendix}
\bigskip

 \bigskip 

In \S 2, we have assumed that the material metric is flat.
 Since
there
is no reason to exclude  non flat material space, we will
consider  the case with non-flat material metric.
It turns out that  the
arguments used to prove the local existence
and global properties in \S 3 and \S 4 applies to the
case of  non flat
material space as well..

Because of the spherical symmetry, the most general form allowed for 
the material metric has the form
$$dy^2 + f^2(y)\left[ d\bar\theta ^2 + 
 {\rm sin}^2 \bar \theta\  d\bar\phi ^2\right]
$$
where $f$ is a non negative smooth function  
such that $f(y)=y \cdot g(y^2)$ for some smooth function $g$ with 
$g(0)=1$.
  We will impose an extra assumption that $f'\ge 0$ and $g$ is a 
bounded function on $\bf R$.
With this material metric, the strain tensor is 
$$S=-{\rm log} {{y'(r)}\over {\sqrt{e^{\ll(r)}}}}\left( e^{\ll(r)}dr^2\right)
-{\rm log} {{f(y)}\over r}\left( r^2 d\theta^2\right)
-{\rm log} {{f(y)}\over r}\left( r^2 {\rm sin}^2 \theta d\phi^2\right)
$$
Note that this expression can be obtained simply by replacing
in (6) $y(r)$ by $ {f(y(r))}$. 
As in  \S 2, we proceed to choose the invariants $u$ and $v$. This time
we choose
$$u:= {\rm log} \left({{r y'}\over {{f(y)}\sqrt {e^\ll}}}\right)
\eqno(48)$$
$$v:= \aa + u = -3 {\rm log} \left( {{f(y)}\over r} \right)
\eqno(49)$$
Then we have the same stress-strain relationship (10) as in the
case of flat material metric. Consequently the expressions in the
 the
Einstein equations and the  Tolman-Oppenheimer-Volkoff
equation (12)-(24) are unchanged. The only deviation from the
flat material  case is that auxiliary equation (for $v'$) is 
different.
$$2xv' = 3-3\sqrt b e^u\cdot
  f'(y)\eqno(50)$$ 
Here we can consider $y$ as an expression written in terms of
$x$ and $v$  since $f$ is invertible near
$y=0$. 

We will show that the statement of 
 Theorem 2 in \S 3 about the local existence of solutions is also true
  for the system
(26),(27),(50).  First, with the  
substitutions $u= xu_1$ and $v=v_o+xv_1$,  we will show
$$x^{-1} \left(1-\sqrt b e^u \cdot f'(y)\right) = -w -u_1 + O(x).\eqno(51)$$
Once this has been
shown,  the local existence result 
for the system (26),(27),(50) follows immediately by following the
proof of Lemma 1. (Recall that
in the proof of Lemma 1 
the auxiliary equation  was used only for the verification of 
(35), which in the current case corresponds to (51).) 
The proof of (51) follows.
From (49), $y= f^{-1}\left(r e^{-{v\over 3}}\right) = O(r)$ as $r\to 0$.
So, by the assumption on $f$, we have $f'(y)= g(y^2) + 2y^2\  g '(y^2)
=1+ O(y^2)= 1+O(r^2)= 1+ O(x)$.  So, the left hand side of (51) 
can be treated as if there were no $f'(y)$  in the second term,
which is the case in the proof of Lemma 1.
 Thus (51) has been verified and the local existence follows.

To obtain the global result stated in Theorem 3  in \S4,
 we  only need  changes 
in the arguments leading to the existence of upper and lower bounds of $v$.
To show that $v$ is bounded from above, in \S 4, we have used the fact that
the second term on the right hand side of (50) is positive.
The same argument applies in the current case since  we have assumed
$f'\ge 0$.  To show that
$v$ is bounded from below, it suffices to show that the second term on the
right hand side of the equation (50) is bounded
 from above.
 The upper bound of $u$ and $b:=e^\ll$ can be
 obtained
independently of the equation (50) (See the proof of Lemma 2 and the 
Step II which follows this lemma.) So, it remains to show that $f'(y)$ is
bounded from above. To show that $f'(y)$ is bounded, it suffices to show that
$y$ is bounded since $f'$ is a continuous function. This will be done
by using the equation (48). Since  we know that $e^\ll$ and $u$ are 
bounded from above, it follows from (48) that $y' \le K f(y)$ for some
constant $K$. But, we have assumed that $f(y)= y \  g(y^2)\le y\cdot C$
for some constant $C$. So, $y'\le K y$, and we have the bound for $y$.

\bigskip\bigskip

\noindent{\bf References}
\medskip
\baselineskip=\normalbaselineskip

{

\parindent 16 pt
\ref {[1]} A. D. Rendall and B. G. Schmidt, 
Existence and Properties of Spherically Symmetric
Static Fluid Bodies with a Given Equation of State, 
{\it Classical and Quantum Gravity } {\bf 8} (1991)
985-1000.

\ref {[2]} T. Makino,
On Spherically Symmetric Stellar Models in 
General Relativity,
{\it Journal of  Math. of Tokyo Univ.} {\bf 38, No. 1} (1998)
55-69.

\ref {[3]} R.L. Bowers and E.P.T. Liang,\ \  
Anisotropic Spheres in General Relativity,\ \ 
{\it Astrophysical Journal} {\bf 188} (1974)
657-665.

\ref {[4]} L. Herrera, G.J. Ruggeri and L. Witten,\ \ 
{\it Astrophysical Journal} {\bf 234} (1979)
1094.

\ref {[5]} G.A. Maugin, {\it J. Math. Phys.} {\bf 19} (1978) 1212.

\ref {[6]} J. Kijowski and G. Magli, \ \ \  
A Generalization  of  Relativistic
Equilibrium  Equations for a
Non-Rotating Star,  {\it General Relativity and Gravitation}
 {\bf vol. 24},   No.2   ( 1992) 139-158.

\ref {[7]} J. Kijowski and G. Magli,
Relativistic Elastomechanics a a Lagrangian Field Theory, 
{\it Journal of Geometry and Physics} 
{\bf 9} (1992)  
207-223.

\ref
{[8]} B. Carter and H. Quintana,
 Foundations of General Relativistic High-Pressure
Elasticity Theory,
{\it Proc. R. Soc. Lond. A.}
{\bf 331} (1972)
57-83.

\ref {[9]} W.C. Hernandez, Jr.,
Elasticity Theory in General Relativity
{\it Physical Review D,} {\bf vol. 1}, No. 4 (1970) 1013-1018.

\ref{[10]} T.W. Baumgarte and A.D. Rendall,
Regularity of Spherically Symmetric Static Solutions of the
Einstein Equations,
{\it Classical and Quantum Gravity}
{\bf 10} (1993) 327-332.

\ref{[11]} G. Rein,
Static Solutions of the Spherically Symmetric Vlasov-Einstein
System,
{\it Math. Proc. Camb. Phil. Soc.  }
{\bf 115 } (1994  ) 559-570.
 
}

\bye